\documentclass[amssymb,aps,twocolumn,floats,showpacs]{revtex4-2}
\usepackage{color,graphicx,pstricks,float}
\usepackage{natbib}

\usepackage{fancybox}
\usepackage{epsfig}
\usepackage{graphicx}
\usepackage{tabularx}
\usepackage{inputenc}
\usepackage{amsmath,braket,amsfonts,amsmath,mathtools,mathrsfs}
\usepackage{amssymb}
\usepackage{gensymb}
\usepackage{grffile}
\usepackage{relsize}
\usepackage{array}
\usepackage{comment}

\usepackage{hyperref}
\usepackage[table,dvipsnames]{xcolor}
\usepackage{gensymb}
\usepackage{grffile}
\usepackage{float}

\hypersetup{
colorlinks=true,
linkcolor = blue,
filecolor = blue,
urlcolor = Blue,
citecolor = blue,
pdftitle = {}
}

\begin{document}

\title{Electric field controlled second-order anomalous Hall effect in altermagnets}

\author{Arnob Mukherjee}
\author{Biplab Sanyal}
\author{Annica M. Black-Schaffer}
\author{Ankita Bhattacharya}
\email{ankita.bhattacharya@physics.uu.se}
\address{Department of Physics and Astronomy, Uppsala University, Box-516, S-75120 Uppsala, Sweden}

\begin{abstract}
Altermagnets are a recently discovered class of compensated magnets with momentum-dependent spin splittings and unusual transport properties, even without a net magnetization. In the presence of combined four-fold rotation and time-reversal ($C_4\mathcal{T}$) symmetry, linear and also second-order, driven by a Berry curvature dipole, anomalous Hall responses are forbidden in any pure $d$-wave altermagnet. Nevertheless, here we find that the nontrivial quantum metric of the occupied Bloch states allows for an electric field induced Berry curvature dipole, which generates a strong and tunable second-order Hall current, enabling it to be switched on or off by simply adjusting the relative orientation between
the symmetry-reducing dc field and the ac probe field. 
Specifically, we investigate the electric field induced second-order anomalous Hall response in a two-dimensional Rashba-coupled hybrid altermagnet that interpolates between $d_{x^2-y^2}$ ($B_{1g}$) and $d_{xy}$ ($B_{2g}$) altermagnet symmetry, motivated by recent proposals for mixed-symmetry states. Crucially, the nonlinear signal is highly sensitive to the underlying symmetry of the altermagnetic order at specific doping levels, offering a purely electrical method to distinguish distinct altermagnetic orders. Our results position hybrid altermagnets as a promising platform for controllable nonlinear transport and spintronic applications.
\end{abstract}

\date{\today}
\maketitle

\noindent
The discovery of altermagnets (AMs) has revealed a new class of magnetic order that bridges ferromagnets (FMs) and antiferromagnets (AFMs), exhibiting momentum-dependent spin splitting but zero net magnetization \cite{krempasky2024, Smejkal2022, Smejkal2022_beyond, tamang2024, song2025, gomonay2024, McClarty2024}.~Unlike conventional (N\'eel) AFMs that preserve time-reversal symmetry ($\mathcal{T}$) through spin-compensated sublattices related by translation or inversion ($\mathcal{P}$), AM breaks global $\mathcal{T}$ while their sublattices are instead connected by crystalline point-group operations, such as rotations (${C}_n$) or mirrors ($\mathcal{M}$) \cite{McClarty2024, Yu2025}. 
 Consequently, AMs have been shown to host exotic electronic features, such as symmetry-protected nodal lines, Weyl planes \cite{Jungwirth2024}, and momentum-locked spin textures \cite{naka2025}, all while suppressing stray magnetic fields thus, making AMs promising candidates for advanced spintronic and topological devices \cite{Bose2024, jungwirth2025}.

In AMs, the Berry curvature (BC) exhibits quadrupolar or higher-order multipolar symmetry in momentum space, resulting in a vanishing momentum-space average and, consequently, the absence of any BC-driven linear anomalous Hall response~\cite{Fang2024, Takahashi2025}, despite  broken $\mathcal{T}$. Consequently, the leading anomalous Hall response may only emerge at second order. Recently, such second-order anomalous Hall effect (SAHE) has attracted considerable interest due to its fundamental connection to the quantum geometry of Bloch states and its potential for promising technological applications \cite{Sodemann2015, Ma2019, Gao2014, Liu2021, Wang2021, gao2023, wang2023, Kaplan2024}.
The primary driver for this SAHE is the dipole moment of the asymmetric distribution of the BC over the occupied state, namely the Berry curvature dipole (BCD)  \cite{Sodemann2015, Kaplan2024}. In addition to requiring broken $\mathcal{P}$, SAHE is subject to strict crystallographic symmetry constraints \cite{Sodemann2015}. In $d$-wave AMs, both $\mathcal{T}$ and four-fold rotational ($C_4$) symmetries are broken individually, but the combined $C_4\mathcal{T}$ is preserved. As a consequence, recent works have shown that the second-order anomalous Hall response is forbidden in AMs and subsequently, only nonlinear Hall responses at the third order have so far been considered \cite{Antonenko2025, Rao2024, Fang2024, Sorn2024, Samokhin2025}.

Through a parallel development, recent theoretical~\cite{Ankita2025} and experimental~\cite{Ye2023,Yang2025electric} works have demonstrated that the SAHE can also occur in systems where the inherent BCD is symmetry forbidden, if instead mediated by the Berry connection polarizability (BCP) or, equivalently, the band-normalized quantum metric (QM), i.e., the real part of the quantum geometric tensor.~\cite{Liu2021, Torma23, sala2024}.  An external dc electric field can induce a finite field-induced BCD through the nontrivial QM of the occupied Bloch bands, making the SAHE present even in systems where the inherent BCD-induced SAHE is absent and, importantly, also  provide greater external tunability. Notably, this electric-field driven SAHE is distinct from the QM-dipole-induced intrinsic (independent of relaxation time) SAHE, reported in $\mathcal{PT}$-symmetric AFM \cite{Liu2021}.

In this work we consider two-dimensional (2D) hybrid AMs. 
While prototypical AMs such as RuO$_2$ \cite{fedchenko2024, plouff2025, guo2024} and MnTe \cite{Yamamoto2025, amin2024, Hariki2024} realize distinct $d_{x^2-y^2}$ ($B_{1g}$) and $d_{xy}$ ($B_{2g}$) spin-splitting, respectively~\cite{fedchenko2024, smejkal2020, McClarty2024}, theoretical classifications have emphasized that these two order parameters, belonging to different irreducible representations of the tetragonal point group $D_{4h}$, can in principle coexist once the crystal symmetry is reduced to one of its lower-symmetry subgroups~\cite{Fernandes2024, Ramires2025}. Moreover, lowering the crystal symmetry, through strain \cite{leon2025, Karetta2025, Khodas2025, li2025} or surface termination \cite{Ramires2025, Samir2025}, may induce mixing between otherwise pure altermagnetic orders. Experimental techniques such as nanoscale domain engineering~\cite{amin2024, Hariki2024} also allow for controlled tuning between these symmetry limits. Motivated by these developments, we consider a hybrid altermagnetic order that continuously interpolates between the $B_{1g}$ and $B_{2g}$ irreducible representations, capturing the possibility of mixed-symmetry states in realistic materials. We also consider AMs that inherently host Rashba spin-orbit coupling (RSOC) due to common structural asymmetry~\cite{Knolle24,Linder24}, which ensures the broken $\mathcal{P}$ required for any finite SAHE~\cite{Sodemann2015}.

Using hybrid AMs with RSOC as realistic platform for AMs, we theoretically establish the existence of an electric field induced BCD-assisted SAHE in AMs. In particular, this exploits the BCP or equivalently, the QM, to overcome the $C_{4}\mathcal{T}$ constraints that otherwise prevents an inherent BCD-induced SAHE. 
Quite remarkably, we also find that the electric field induced SAHE varies substantially in magnitude between the $d_{x^{2}-y^{2}}$ and $d_{xy}$ limits in certain doping regimes, thus enabling a purely electrical route to distinguish the two altermagnetic orders. Our findings establish AMs as a powerful platform to realize highly tunable nonlinear Hall transport already at second order, driven by quantum geometric effects.

\noindent
\textit{Model hybrid altermagnet (AM).}--- To investigate the SAHE in hybrid AMs, we introduce a minimal two-band Hamiltonian defined on a 2D square lattice. In this model, we phenomenologically mix the $B_{1g}$ and $B_{2g}$ components of the altermagnetic orderings to capture the essential physics of a hybrid AM. The tight-binding Hamiltonian is given by \cite{Farajollahpour2025, Tanaka2024, Ghorashi2024}
\begin{equation}
H(\mathbf{k}) = \epsilon(\mathbf{k})\,\sigma_0 + \mathbf{h}(\mathbf{k})\cdot\boldsymbol{\sigma},
\label{eq:Hamiltonian}
\end{equation}
where $\sigma_{\nu}$ denote the Pauli matrices for $\nu = 1, 2, 3$ and the
$2 \times 2$ identity matrix for $\nu = 0$ acting in the spin-sublattice subspace. The components of the Hamiltonian are
\begin{align}
\epsilon(\mathbf{k}) &= -2t\,(\cos k_x + \cos k_y) - \mu, \\
\mathbf{h}(\mathbf{k}) & =\{-\lambda \,\sin k_y , \, \lambda \sin k_x , \, t_{\text{am}}\,\alpha \,(\cos k_x - \cos k_y)  \nonumber \\
 & + t_{\text{am}}\,(1 - \alpha) \,\sin k_x \sin k_y \}
\end{align}
where $t$ denotes the nearest-neighbor hopping amplitude, and $\mu$ is the chemical potential, $\lambda$ denotes the RSOC strength, and $t_{\text{am}}$ is the strength of the altermagnetic spin splitting . We parameterize the relative weight of the two altermagnetic orders by $0 \leq \alpha \leq 1$. In two dimensions, AMs inherently exhibit RSOC due to structural asymmetry~\cite{Knolle24,Linder24}. 
The case $\alpha=1$ corresponds to a pure $d_{x^2-y^2}$ AM order arising from nearest-neighbor spin-dependent hopping, while $\alpha=0$ yields pure $d_{xy}$ AM order governed by next-nearest-neighbor spin-dependent hopping. Intermediate values of $\alpha$ thus describe a hybrid AM that seamlessly interpolates between the two irreducible representations $B_{1g}$ and $B_{2g}$ of the point group $D_{4h}$. Such hybridization significantly enriches the spin splitting .

The square lattice Hamiltonian in Eq.~(\ref{eq:Hamiltonian}) may be relevant for tetragonal AMs such as RuO$_2$ \cite{fedchenko2024, plouff2025, guo2024} and MnTe \cite{Yamamoto2025, amin2024, Hariki2024}, and is directly relevant to quasi-2D layered compounds like Ca$_3$Ru$_2$O$_7$ that host both altermagnetism and RSOC \cite{leon2025}. Additionally, it naturally accommodates the $d_{x^2-y^2}$ ($B_{1g}$) and $d_{xy}$ ($B_{2g}$) \cite{Smejkal2022, Smejkal2022_beyond} orders whose hybridization we study.

\textit{Second-order anomalous Hall effect (SAHE).}--- Although the Hamiltonian in Eq.~(\ref{eq:Hamiltonian}) breaks $\mathcal{T}$, it exhibits no anomalous Hall response at linear order in an applied ac electric field  $\mathcal{E}^\omega$ with frequency $\omega$, owing to the presence of combined $C_4 \mathcal{T}$ symmetry. Consequently, the leading anomalous Hall transport may at most arise at second order in $\mathcal{E}^\omega$. Within semi-classical Boltzmann theory, the second-order anomalous Hall current can be expressed as $j_i^{\text{AH}} = \chi^{\text{AH}}_{i j k}\,\mathcal{E}^\omega_j\,\mathcal{E}^\omega_k$, with the second-order transverse conductivity tensor $\chi^{\text{AH}}_{i j k}$~\cite{Sodemann2015}
\begin{align}
   \chi^{\text{AH}}_{i j k} = \epsilon_{ilk}\, \frac{e^3 \tau}{2\hbar^2 (1 + i \omega \tau)}\, \sum_{n}\, \int \frac{d^d \mathbf{k}}{(2 \pi)^d}\, f_0 \,(\partial_{k_j} \Omega^n_{ l}),
    \label{chi_def}
\end{align}
where $\epsilon_{ilk}$ is the Levi-Civita symbol and 
$f_0$ is the equilibrium Fermi-Dirac distribution function and $\Omega^n_{l}$ is the $l$-th component of BC of the $n$th band. Here, $\tau$ is the relaxation time that makes the SAHE an extrinsic effect. The integral is the first moment of the BC over the occupied states, known as the Berry curvature dipole (BCD)~\cite{Sodemann2015}
\begin{align}
  D_{j l} =   \sum_{n}\, \int \frac{d^d \mathbf{k}}{(2 \pi)^d}\, f_0 \,(\partial_{k_j} \Omega^n_{l}).
    \label{BCD_def}
\end{align}
For the BCD, or equivalently $\chi^{\text{AH}}_{i j k}$, to be nonzero, inversion symmetry $\mathcal{P}$ must be broken, otherwise, the momentum derivative $\partial_{k_j} \Omega^n_{l} (\mathbf{k})$ transforms as an odd function of momentum, causing its Brillouin zone integral to vanish. In Eq.~(\ref{eq:Hamiltonian}), the Rashba term $\lambda $ explicitly breaks $\mathcal{P}$. Additionally, the BCD obeys a severe symmetry constraint, namely the presence of two or more mirror lines in a crystal forces the SAHE to still vanish, while a single mirror line forces it to be orthogonal to that mirror plane \cite{Sodemann2015,ortix2021,Nandy2019}. Thus, mirror symmetries $\mathcal{M}_{x,y}$, when present, constrain BCD and thus, the SAHE.

Pure $d_{x^2 - y^2}$ altermagnetic order, i.e., when $\alpha=1$ in Eq.~(\ref{eq:Hamiltonian}), breaks both $\mathcal{M}_{x}$ and $\mathcal{M}_{y}$ mirror symmetries, while purely $d_{xy}$, i.e., $\alpha=0$, preserves both $\mathcal{M}_{x,y}$. The presence of both $\mathcal{M}_{x,y}$ thus directly prohibits any BCD-induced SAHE for $\alpha=0$. In contrast, the absence of both mirror planes for $\alpha=1$, in principle, should allow SAHE. However, the presence of combined $C_4 \mathcal{T}$ symmetry in AMs has been shown to completely forbid BCD-driven SAHE for any values of $\alpha$, thus forcing the leading order Hall transport to be third order~\cite{Sankar2024, Xiang2023, Fang2024, Takahashi2025}. 
Next we will show that the nontrivial QM of the Bloch states actually still enables an extrinsic second-order transverse response in the presence of a dc electric field in AMs. 

\begin{figure}[t!]
    \centering
    \includegraphics[width=\columnwidth]{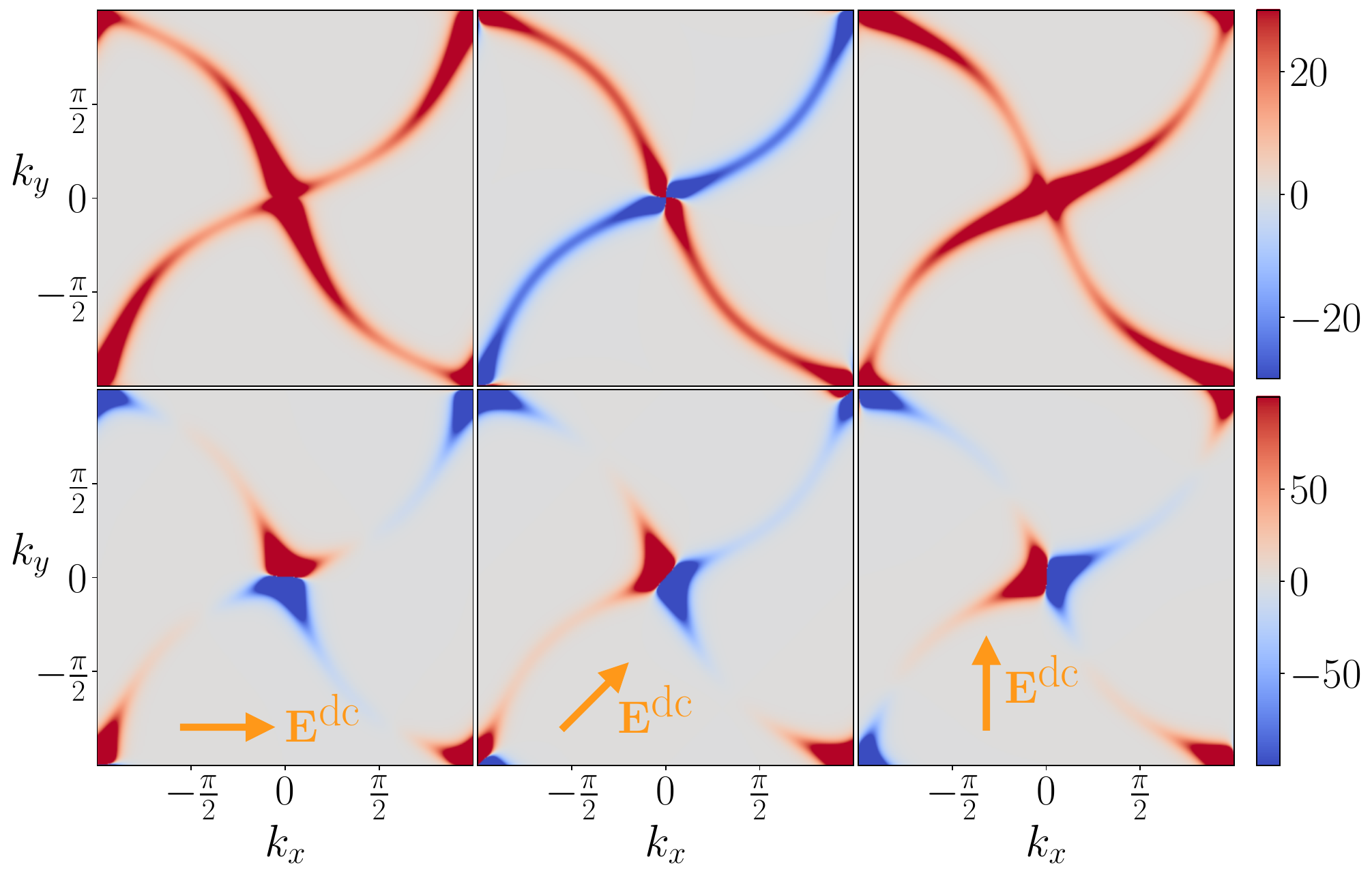}
    \caption{Berry connection polarizability (BCP) tensor components for one of the band: (a) $G^{1}_{xx}$, (b) $G^{1}_{xy}$, and (c) $G^{1}_{yy}$ in the first Brillouin zone.
Field-induced Berry curvature for the same band , $\Omega^{\textrm{E}}(\mathbf{k})$, for dc-field orientations (d) $\theta=0$, (e) $\pi/4$, and (f) $\pi/2$, with range arrows indicating electric field directions. Calculations performed with $t=1.0$, $t_{\text{am}}=0.5 t$, $\alpha=0.5$, $\mu=0.3t$, and $\lambda=0.08t$.}
    \label{fig:fig1}
\end{figure}

\noindent
{\textit{Electric field induced SAHE.}}---
 In the presence of a symmetry-reducing dc electromagnetic field $\mathbf{E}^\text{dc}$, the BC of the Bloch electrons is modified due to the positional shift of Bloch electrons arising from interband mixing~\cite{Gao2014}.  
 This results in a field-induced Berry connection that is expressed as $\mathcal{A}^{E} = \stackrel{\leftrightarrow}{\mathbf{G}}\, \mathbf{E}^\text{dc}$, where $\stackrel{\leftrightarrow}{\mathbf{G}}$ is a second-rank tensor, called the BCP. The component  $ G_{ab}^n$ of the BCP  tensor is defined as \cite{Gao2014,Lai2021,Liu2022}
\begin{equation}
    G_{ab}^n ({\bf k}) = 2~\text{Re}~\sum_{m \neq n} \frac{\mathcal{A}_a^{nm}({\bf k})~\mathcal{A}_b^{mn}({\bf k})}{\epsilon_n({\bf k}) - \epsilon_m({\bf k})}.
    \label{BCP-def}
\end{equation}
Here, the interband Berry connection is $\mathcal{A}_a^{nm}({\bf k}) = \langle u_n | i \partial_{a} | u_m \rangle$, where $\partial_a \equiv \partial_{k_a}$ and $|u_n \rangle$ is the periodic part of the $n$-th unperturbed Bloch state, with $\epsilon_n$ the corresponding band energy. Notably, $ G_{ab}^n ({\bf k})$ is closely linked with the QM, $g_{ab}^n = 2 \;\sum_{m \neq n} \text{Re} [\mathcal{A}_a^{nm}({\bf k}) \; \mathcal{A}_b^{mn}({\bf k})]$, the real part of the quantum geometric tensor \cite{Torma23}.
A nontrivial BCP, or equivalently QM, generates in turn a field-induced BC $\mathbf{\Omega}^E_n = \nabla_k \times \mathcal{A}^E_n$, which, similar to the original BC of the Bloch band, acts like a pseudo-magnetic field in ${\bf k}$-space 
which yields a SAHE, just as the original BC in Eq.~(\ref{chi_def}). The field-induced BC is especially important in those systems where either the intrinsic BC vanishes or the original BCD-induced SAHE is forbidden due to the symmetry constraints, as discussed above for AMs.

In typical Hall transport measurements, the applied electric field and the measured current are both restricted to the $x-y$ plane. Then, an external electric field applied at an angle $\theta$ relative to the $x$-axis, such that ${\bf E}^{\text{dc}} = E^{\text{dc}} (\cos \theta, \sin \theta)$, induces a field-induced BC of the form \cite{Lai2021, Pal2024,Ankita2025}, 
\begin{align} \label{eq.E_indc_BC}
  \Omega^{\text{E}}_{n z} &=E^{\text{dc}}[(\partial_{k_x} G^n_{yx}- \partial_{k_y} G^n_{xx})\,\cos\theta \\ \nonumber
     & + (\partial_{k_x} G^n_{yy}- \partial_{k_y} G^n_{xy})\,\sin\theta],
\end{align}
which is dependent on both the magnitude and direction of $\mathbf{E}^\text{dc}$. This field-induced BC can lead to a finite field-induced BCD following Eq.~(\ref{BCD_def}) with an angular dependence ${\bf D}^{\text{E}} (\theta) = \big(D_{xz}^{\text{E}}(\theta), D_{yz}^{\text{E}}(\theta)\big)$. Due to the lowered symmetry in the presence of ${\bf E}^{\text{dc}}$, a field-induced BCD then gives rise to the SAHE. 

Finally, plugging the angular dependence of the BCD into Eq.~\eqref{chi_def}, and recasting it through $\chi^{\text{AH}}= j^{2\omega}/(E^\omega)^2$, we arrive at the second-harmonic current $j^{2\omega}$~\cite{Ye2023,Ankita2025}  
\begin{align}
    j^{2\omega}= -\frac{e^3 \tau}{2 (1 + i \omega \tau) \hbar^2}\, (\hat{z}\times \mathbf{E}^\omega ) \,[\mathbf{D}^{\text{E}}(\theta) \cdot \mathbf{E}^\omega],
    \label{acHall2nd}
\end{align}
where $\mathbf{E}^\omega$ is the in-plane ac Hall probe field that makes an angle $\phi$ with the $x$-axis and satisfies the condition $E^\omega \ll E^{\text{dc}}$.
As seen from Eq.~(\ref{acHall2nd}), the maximum Hall response is obtained when $\mathbf{E}^\omega \parallel \mathbf{D}^{\text{E}}(\theta)$, while it vanishes for $\mathbf{E}^\omega \perp \mathbf{D}^{\text{E}}(\theta)$. Overall, the angular dependence $j^{2\omega}$ is intricate as it not only varies with $\theta$ through $\mathbf{D}^E(\theta)$ but also with $\phi$, thus providing a high degree of external tunability of the SAHE. In $2$D, only the two components $\chi_{yxx}$ and $\chi_{xyy}$ of the tensor $\chi^{\text{AH}}_{ijk}$ are finite and independent. Thus $\chi^{\text{AH}}$ simplifies to \cite{Liu2021, Ankita2025}

\begin{align}
    \chi^\text{AH} (\theta, \phi)= \chi_{yxx} (\theta) \cos{\phi}-\chi_{xyy}(\theta) \sin{\phi}.
    \label{chi_theta_phi}
\end{align}
\begin{figure}[b!]
    \centering
    \raisebox{1ex}{\includegraphics[width=\columnwidth]{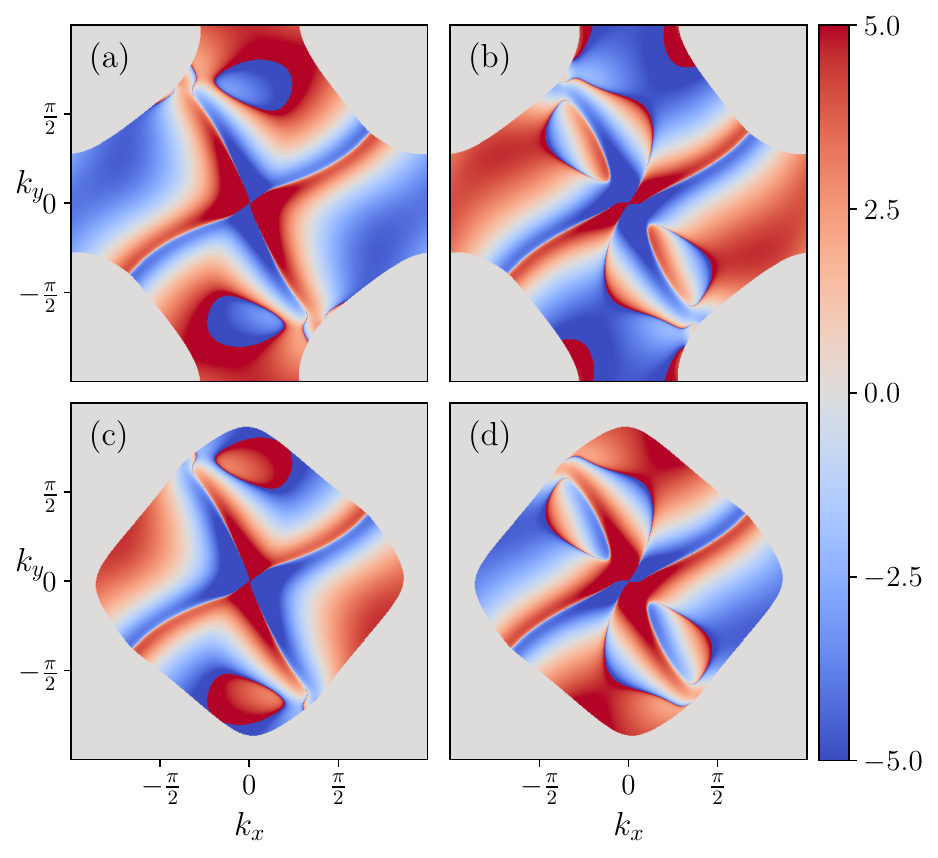}}
    \caption{Momentum distribution of the Fermi-function-weighted derivatives of the field-induced Berry curvature in log-scale for the two bands, (a) $f_0~\partial_{k_x}\Omega^{\mathrm{E}}_1$, (b) $f_0~\partial_{k_y}\Omega^{\mathrm{E}}_1$, and for band 2, (c) $f_0~\partial_{k_x}\Omega^{\mathrm{E}}_2$, (d) $f_0~\partial_{k_y}\Omega^{\mathrm{E}}_2$, evaluated at $\theta = \pi/4$. Parameters are the same as in Fig. \ref{fig:fig1}.}
    \label{fig:fig2}
\end{figure}

\noindent
\textit{SAHE in altermagnets.}--- To illustrate how the electric field induced SAHE emerges in AMs described by Eq.~(\ref{eq:Hamiltonian}), we start by showcasing the $\mathbf{k}$-resolved distribution of the components of the BCP tensor, or equivalently, band-normalized QM for one of the two bands, setting $\alpha=0.5$ in Fig.~\ref{fig:fig1} (a-c). For the other band, the contributions are exactly equal in magnitude but opposite in sign.  For pure altermagnetic orders, i.e., for $\alpha=0$ or $\alpha=1$, the components of BCP tensors are shown in the Supplementary Material (SM) \cite{SM}.
The corresponding band structures for these three $\alpha$ values are shown in the SM \cite{SM}. Consistent with expectations, the BCP values exhibit maxima at the band-touching points, due to the denominator in Eq.~\eqref{BCP-def}. 
The presence of a finite BCP leads to a finite BC, in accordance with Eq.~(\ref{eq.E_indc_BC}), when an external dc electric field $\mathbf{E}^\text{dc}$ is applied. For $\mathbf{E}^\text{dc}$ oriented along the $x$ ($\theta=0$), $x=y$ ($\theta=\pi/4$), and $y$($\theta=\pi/2$), the field-induced BC $\Omega^{E}$ are shown in Fig.~\ref{fig:fig1} (d)-(f), respectively. The $\mathbf{k}$-resolved distribution of $\Omega^{E}$ shows a clear dipolar structure. Not only the magnitude, but also the direction of the $\Omega^{E}$ and, thus, also the resulting BCD, can be externally tuned by changing the direction of $\mathbf{E}^\text{dc}$~\cite{Ankita2025}.

\begin{figure}[t!]
    \centering
    \raisebox{1ex}{\includegraphics[width=0.95\columnwidth]{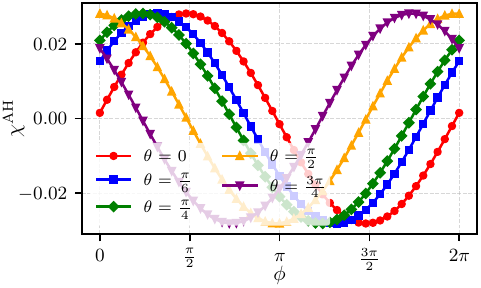}}
    \caption{Second-order Hall conductivity $\chi^{\mathrm{AH}}$ as a function of angle $\phi$, set by the ac driving field $\mathbf{E}^\omega$, for
several different angles $\theta$, set by the symmetry-breaking static field $\mathbf{\mathbf{E}}^{\mathrm{dc}}$, with the $x$-axis. $\chi^{\mathrm{AH}}$ is scaled by the prefactor $\mathcal{A}=\frac{e^3 \tau}{2 (1 + i \omega \tau) \hbar^2} $  and the parameters are the same as in Fig.~\ref{fig:fig1}.}
    \label{fig3}
\end{figure}

Once a finite and field-tunable BCD is confirmed, it is straightforward to express $\chi^\text{AH}$ in terms of $D_{xz}$ and $D_{yz}$ following Eq.~(\ref{chi_def}-\ref{BCD_def}).
Figure~\ref{fig:fig2} shows the momentum-resolved components of the Fermi-weighted Berry curvature dipole, i.e., the integrand of Eq.~(\ref{BCD_def}) for both bands again for 
 $\alpha = 0.5$, with $\mathbf{E}^\text{dc}$ oriented along the $x = y$ direction. These are then integrated over the Brillouin zone and summed over bands to yield the finite field-induced BCD. Finally, following Eq.~(\ref{chi_def}), with it angular dependence in Eq.~(\ref{chi_theta_phi}), we find an overall nonzero second-order Hall response. 
The complex angular variation of the resulting second-order Hall conductivity $\chi^\text{AH} (\theta, \phi)$ in a hybrid AM is depicted in Fig.~\ref{fig3}. It offers significant experimental control over the second-order Hall current, enabling it to be switched on or off simply by adjusting the relative orientation of $\mathbf{E}^{\text{dc}}$ and $\mathbf{E}^\omega$. These behaviors are similarly observed for the pure altermagnetic order, i.e., for $\alpha=1$ and $\alpha=0$, see the SM \cite{SM}.

\begin{figure}[t!]
    \centering
    \raisebox{1ex}{\includegraphics[width=\columnwidth]{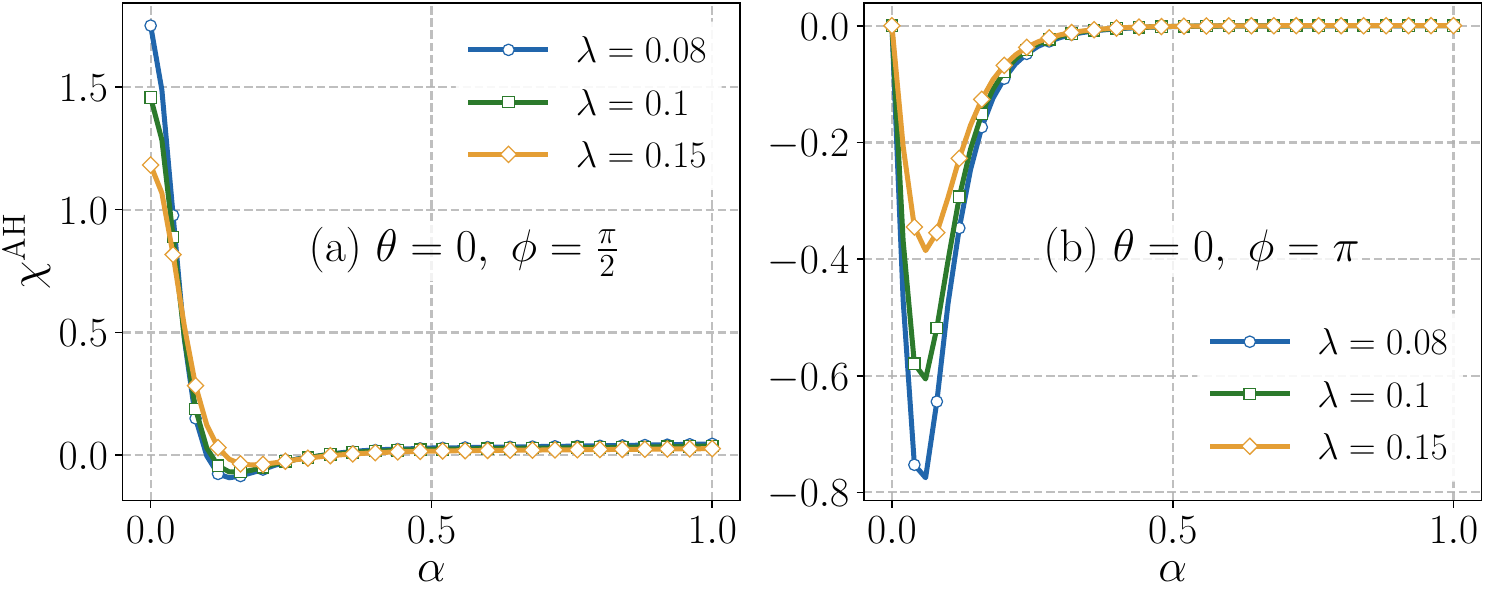}}
    \caption{Second-order Hall conductivity $\chi^{\mathrm{AH}}$ as a function of $\alpha$ for different RSOC $\lambda$ values and field orientations: (a) $\theta=0$, $\phi=\pi/2$ and (b) $\theta=0$, $\phi=\pi$ with $t = 1$, $t_{\text{am}}=0.5 t$, and $\mu=0.3t$ (same as Fig.~\ref{fig:fig1}). }
    \label{fig4}
\end{figure}

Remarkably, we find that the electric field induced SAHE exhibits a strong sensitivity in magnitude between different altermagnetic orders, at least for a range of doping levels, set by $\mu$. The response is maximized at $\alpha=0$, corresponding to a pure $d_{xy}$-type altermagnetic order when $\mathbf{E}^\text{dc} \perp \mathbf{E}^\omega$, but as $\alpha$ increases, the response $\chi^{\text{AH}}$ decreases and becomes negligible for $\alpha \rightarrow 1$, i.e., pure $d_{x^2-y^2}$ order, see Fig.~\ref{fig4}(a). In contrast, when $\mathbf{E}^\text{dc} \parallel \mathbf{E}^\omega$, $\chi^\text{AH}$ is found to be vanishingly small for both $\alpha=0$ and $\alpha=1$ as shown in Fig.~\ref{fig4}(b). Thus, simply tuning the direction of either $\mathbf{E}^\text{dc}$ or $\mathbf{E}^\omega$, provides a straightforward way to conclusively distinguish between pure orders $d_{xy}$ and $d_{x^2-y^2}$. For the parameter choices used in the presented results, this doping range is found to be $-t< \mu < t$. We have also tested our results for smaller and larger $t_{\text{am}}$, which shifts the corresponding doping range, although it is still clearly present. In this doping regime, the system exhibits two hole-like Fermi pockets for $\alpha = 0$, whereas for $ \alpha=0.5$ and $ \alpha=1$, it features one electron and one hole Fermi pocket, see SM \cite{SM}. When the Fermi surfaces (FSs) have the opposite curvature, their contributions to Hall transport tend to cancel each other, resulting in a nearly vanishing response. In contrast, when both FSs have the same curvature, their contributions add constructively, leading to a finite overall response. For example, in the doping regime with two electron pockets (regardless of $\alpha$), the field-induced SAHE remains nonzero for both pure and hybrid altermagnetic symmetries, see the SM \cite{SM}. In this scenario, pure altermagnetic orders cannot be distinguished based on the second-order response, as all yield a finite signal.

These findings unambiguously establish the presence of a nonlinear Hall response in AM at second order and also offer an all-electrical approach to resolve pure altermagnetic orderings. As shown in Fig.~\ref{fig4}, the electric field induced SAHE increases when the strength of RSOC $\lambda$ decreases. Still, we recall that no SAHE is possible for $\lambda=0$ due to preserved $\mathcal{P}$. At first glance, this may seem non-intuitive. However, we note that for finite $\lambda$, RSOC opens gaps at the band-touching points in altermagnetic metals. When $\lambda$ is small, these gaps are narrow, leading to an enhanced BCP (see Eq.~(\ref{BCP-def})), which in turn amplifies the field-induced SAHE. The evolution of the two spin-split FSs with $\lambda$ in various doping limits are shown in the SM \cite{SM}.

In summary, we demonstrate that Rashba-coupled hybrid AMs provide a viable platform for electric field induced SAHE that arises due to the nontrivial QM of the Bloch bands. Although combined $C_{4}\mathcal{T}$ symmetry forbids both the linear anomalous Hall effect and the usual BCD-driven second-order signal, an external dc field lowers the symmetry and generates a finite field-induced BCD via the nontrivial QM of the occupied Bloch states. The intricate angular variation of the resulting second-order Hall conductivity $\chi^\text{AH} (\theta, \phi)$ provides substantial experimental control over the second-order Hall current, allowing it to be toggled on or off by merely adjusting the relative orientation between the external dc field $\mathbf{E}^{\text{dc}}$ and ac probe field $\mathbf{E}^\omega$. In certain doping regimes, its magnitude can even probe the underlying altermagnetic form factor, peaking in the $d_{xy}$ ($B_{2g}$) limit and decreasing toward $d_{x^2-y^2}$ ($B_{1g}$) configuration for $\mathbf{E}^{\mathrm{dc}}\perp \mathbf{E}^\omega$, which then provides a purely electrical means to distinguish the two.

These predictions are directly testable in candidate material platforms such as epitaxial RuO$_2$ \cite{fedchenko2024, plouff2025, guo2024} and MnTe \cite{Yamamoto2025, amin2024, Hariki2024} thin films, where strain or surface termination 
can realize hybrid $B_{1g}$–$B_{2g}$ textures and provide the required inversion-symmetry breaking at interfaces that generate interfacial RSOC. To observe the SAHE, an experimental setup similar to the one used in Refs.~\cite{Ye2023,Yang2025electric} could be employed.
More broadly, our results demonstrate that quantum geometry can enable nonlinear responses in systems where symmetry otherwise suppresses intrinsic BC effects, enabling functionalities such as on-chip frequency doubling and rectification in zero-moment magnets.  
\noindent

A.B and A.M.B.-S. acknowledge financial support from the Swedish Research Council
(Vetenskapsr\aa det) grant no.~2022-03963 and the European Union through the European Research Council (ERC) under the European Union’s Horizon 2020 research and innovation programme (ERC-2022-CoG, Grant agreement no.~101087096). Views and opinions expressed are however those of the author(s) only and do not necessarily reflect those of the European Union or the European Research Council Executive Agency. Neither the European Union nor the granting authority can be held responsible for them.
B.S. acknowledges financial support from Swedish Research Council (grant no.~2022-04309), STINT Mobility Grant for Internationalization (grant no.~MG2022-9386) and DST-SPARC, India (Ref.~No.~SPARC/2019-2020/P1879/SL). A.M. acknowledges the computational resources provided by the National Academic Infrastructure for Supercomputing in Sweden (NAISS) at NSC and PDC (NAISS 2024/3-40) partially funded by the Swedish Research Council through grant agreement no.~2022-06725 and at UPPMAX (NAISS 2025/2-203). B.S. also acknowledges the allocation of supercomputing hours granted by the EuroHPC JU Development Access call in LUMI-C supercomputer (grant no.~EHPC-DEV-2024D04-071) in Finland.

%

\newpage
\pagebreak

\setcounter{equation}{0}
\setcounter{figure}{0}

\renewcommand{\thefigure}{S\arabic{figure}}
\renewcommand{\theequation}{S\arabic{equation}} 

\onecolumngrid
\vspace{10pt}

\begin{center}
{\bf {\Large{Supplemental Material for ``Electric-field-controlled second-order anomalous Hall effect in altermagnets"}}}
\end{center}
In this supplementary material, we present results for the pure altermagnetic orders. We also examine a doping regime in which both pure and hybrid altermagnetic orders feature Fermi pockets with similar curvature, allowing for a direct comparison with the main text, where the results correspond to a doping range in which the altermagnetic orders exhibit either Fermi surfaces with the same curvature or with opposite curvature.

\section{Field-induced Second-order Hall response for pure altermagnetic orders}
\begin{figure}[h!]
    \centering
    \includegraphics[width=0.75\columnwidth]{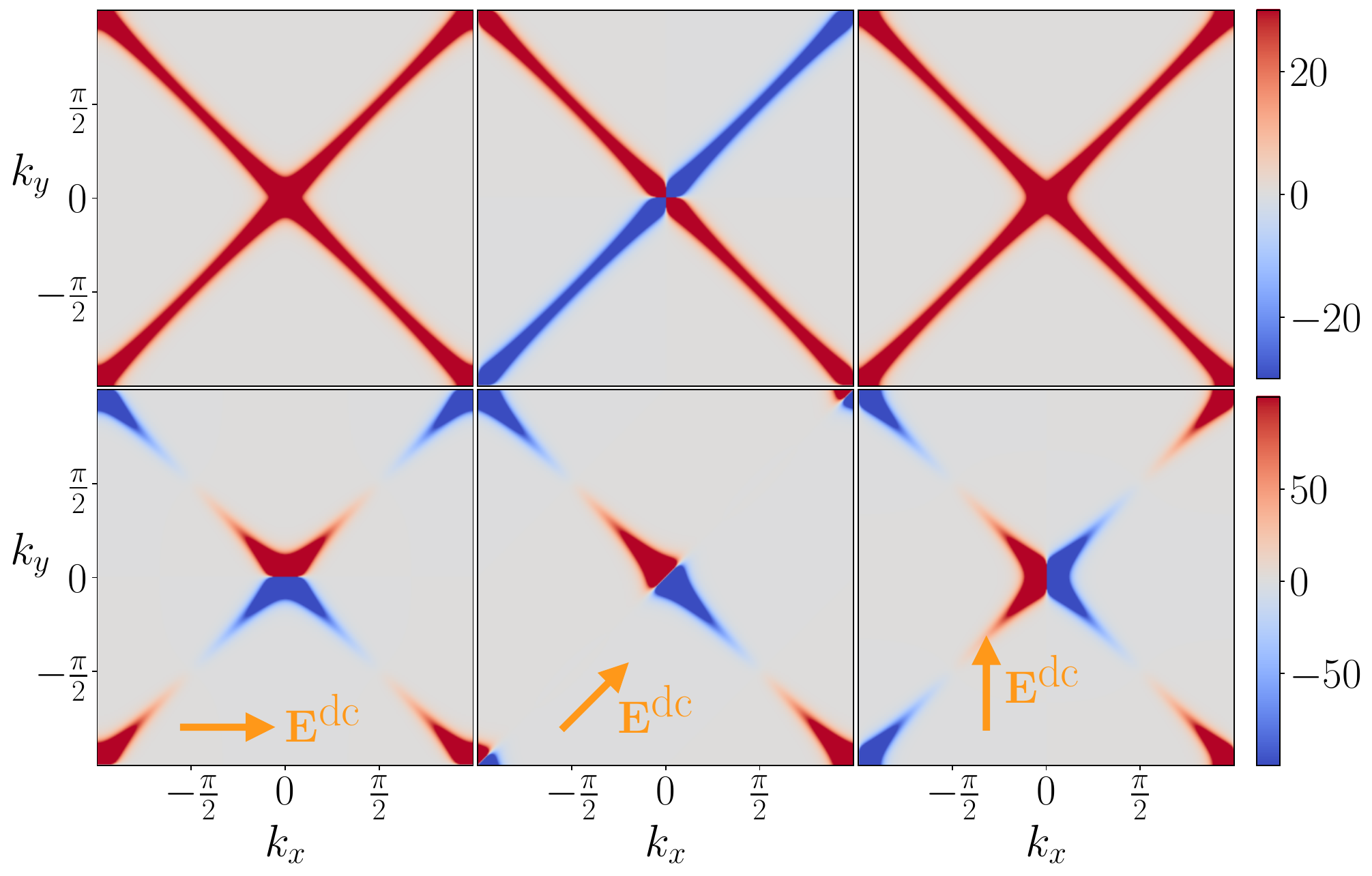} 
  \caption{Berry connection polarizability (BCP) tensor components for one of the band : (a) $G^{1}_{xx}$, (b) $G^{1}_{xy}$, and (c) $G^{1}_{yy}$ in the first Brillouin zone.
Field-induced Berry curvature for the same band , $\Omega^{\textrm{E}}(\mathbf{k})$, for dc-field orientations (d) $\theta=0$, (e) $\pi/4$, and (f) $\pi/2$, with orange arrows indicating electric field directions. Calculations performed with $t=1.0$, $t_{\text{am}}=0.5 t$, $\alpha=1.0$, $\mu=0.3t$, and $\lambda=0.08t$.}
    \label{fig:SM_Fig1a}
\end{figure}

\begin{figure}[t!]
    \centering
    \includegraphics[width=0.65\columnwidth]{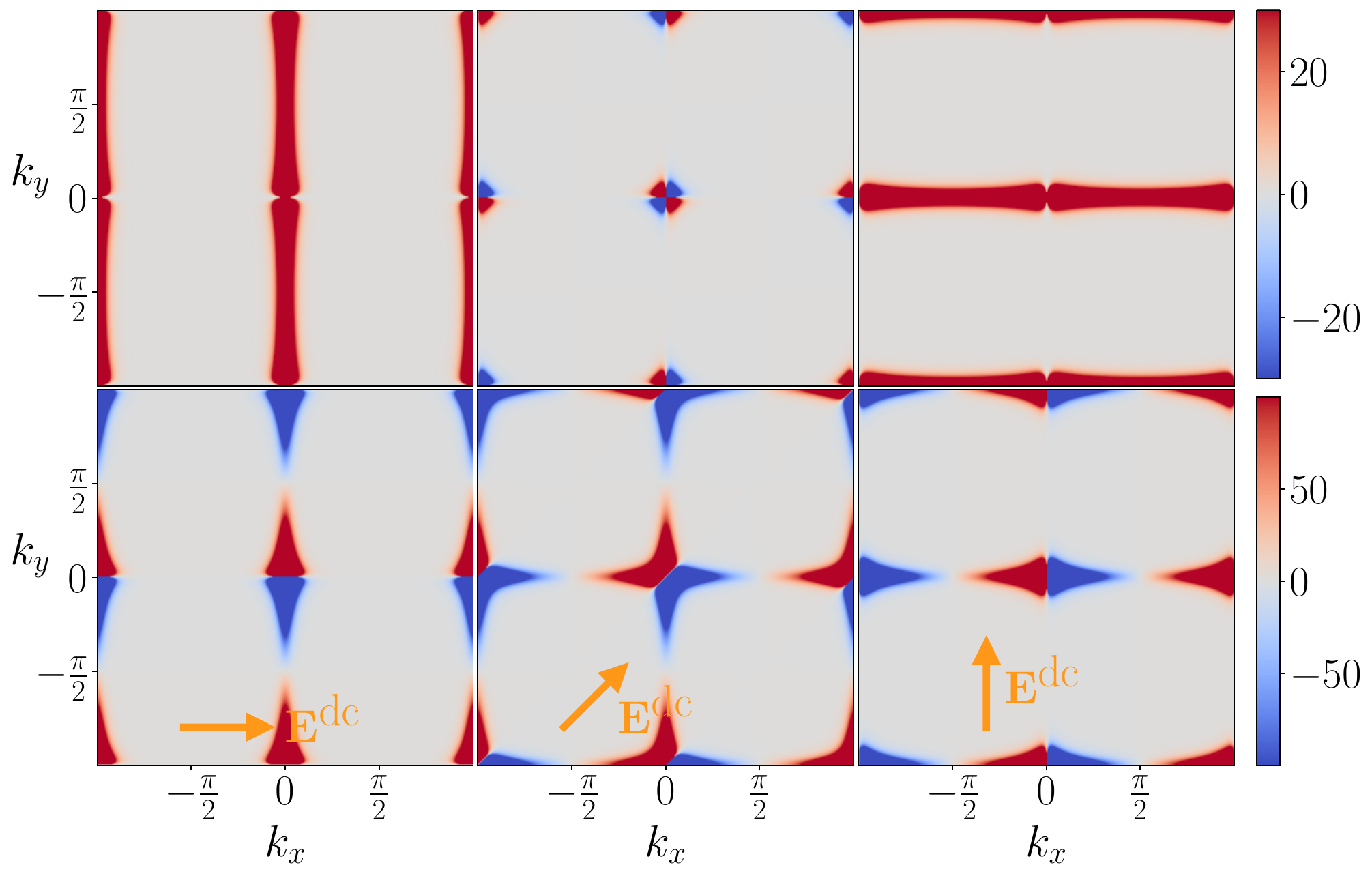} 
\caption{Berry connection polarizability (BCP) tensor components for one of the band : (a) $G^{1}_{xx}$, (b) $G^{1}_{xy}$, and (c) $G^{1}_{yy}$ in the first Brillouin zone.
Field-induced Berry curvature for the same band , $\Omega^{\textrm{E}}(\mathbf{k})$, for dc-field orientations (d) $\theta=0$, (e) $\pi/4$, and (f) $\pi/2$, with orange arrows indicating electric field directions. Calculations performed with $t=1.0$, $t_{\text{am}}=0.5 t$, $\alpha=0.0$, $\mu=0.3t$, and $\lambda=0.08t$.}
    \label{fig:SM_Fig1b}
\end{figure}

In this section we show the electric field induced second-order Hall response for pure altermagnetic orders. For our model Hamiltonian Eq.~(\ref{eq:Hamiltonian}) in the main text we consider hybrid altermagnetic order, i.e., a combination of $d_{x^2-y^2}$ and $d_{xy}$ symmetries by keeping $\alpha=0.5$. Here we instead consider the results for pure $d_{x^2-y^2}$  and $d_{xy}$ altermagnetic order. In Fig.~\ref{fig:SM_Fig1a} and Fig.~\ref{fig:SM_Fig1b}, we show the components of the Berry connection polarizability tensor (BCP) for one of the two bands and the field-induced Berry curvature for three different orientations of the symmetry-breaking field $\mathbf{E}^\text{dc}$ for $\alpha=1$, i.e., for pure $d_{x^2-y^2}$ symmetry and for $\alpha=0$, i.e., for pure $d_{xy}$ symmetry, respectively. These figures are equivalent to Fig.~\ref{fig:fig1} in the main text.

The $\mathbf{k}$-resolved distribution of $\Omega^{E}$ shows a clear dipolar structure, which corresponds directly to a finite field-induced BCD. Not only the magnitude but also the direction of $\Omega^{E}$ and, thus, the resulting BCD can be externally tuned by changing the direction of $\mathbf{E}^\text{dc}$. The corresponding momentum-resolved components of the Fermi-weighted dipole of BC for the two bands for $\alpha = 1$ and $\alpha = 0$, with $\mathbf{E}^\text{dc}$ oriented along the $x = y$ direction, are shown in Fig.~\ref{fig:SM_Fig2a} and Fig.~\ref{fig:SM_Fig2b}, respectively, similar to Fig.~\ref{fig:fig2} in the main text. For $\alpha=1$, the system possesses one electron pocket and one hole pocket, similar to $\alpha=0.5$, while for $\alpha=0$, it has two hole pockets. Figure \ref{fig:SM_FigS3} shows the resulting second-order conductivity $\chi^\text{AH}(\theta, \phi)$ as a function of $\phi$ for various values of $\theta$ for the two pure orders. The second-order response $\chi^\text{AH}(\theta, \phi)$ for $\alpha=0$ is more than an order of magnitude higher than that for $\alpha=1$. When the Fermi surfaces (FSs) are of opposite types, as in the case for $\alpha=0.5$ or $1$, their contributions tend to cancel each other, resulting in a small response. In contrast, when both FSs have same types of curvature as in the case for $\alpha=1$, their contributions add constructively, leading to a higher overall response.

\begin{figure}[h!]
    \centering
    \includegraphics[width=0.85\columnwidth]{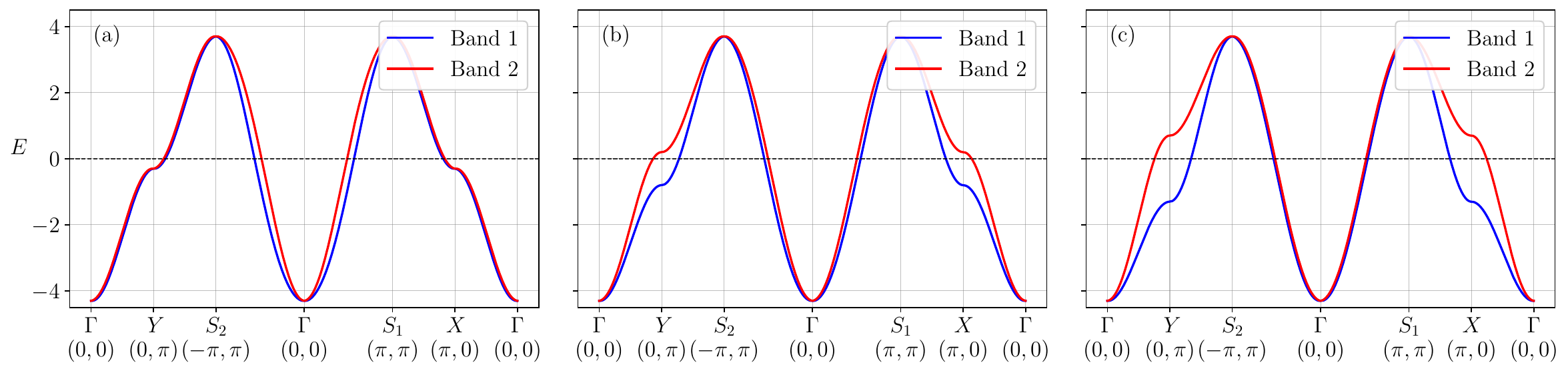} 
   \caption{Band structure along high-symmetry paths for (a) $\alpha$=0 (pure $d_{xy}$), (b) $\alpha$=0.5 (hybrid), and (c) $\alpha$=1 (pure $d_{x^2-y^2}$). 
   Blue and red lines represent the two spin-split bands band 1 and band 2, respectively. The dashed line indicates the Fermi level. Parameters: $t=1.0$, $\mu=0.3 t$, $\lambda=0.08t$, $t_{\text{am}}=0.5t$.}
    \label{fig:SM_FigBS}
\end{figure}
\begin{figure}[t!]
    \centering
    \includegraphics[width=0.65\columnwidth]{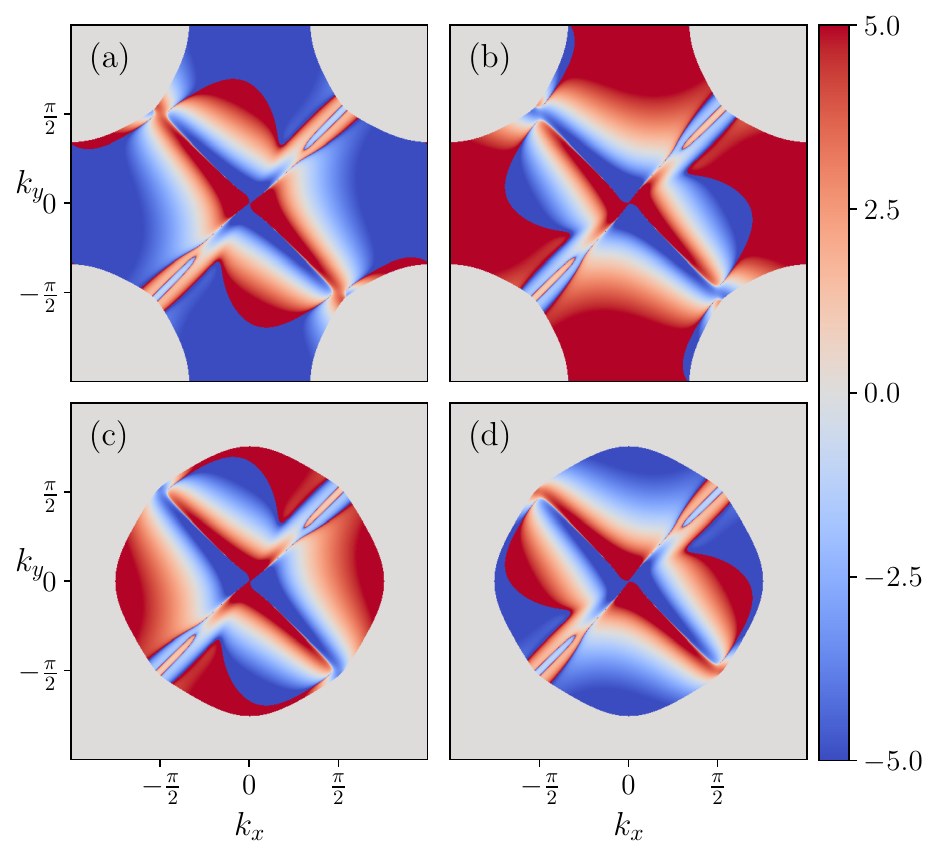} 
    \caption{Momentum distribution of the Fermi-function-weighted derivatives of the field-induced Berry curvature in log-scale for the two bands, (a) $f_0~\partial_{k_x}\Omega^{\mathrm{E}}_1$, (b) $f_0~\partial_{k_y}\Omega^{\mathrm{E}}_1$, and for band 2, (c) $f_0~\partial_{k_x}\Omega^{\mathrm{E}}_2$, (d) $f_0~\partial_{k_y}\Omega^{\mathrm{E}}_2$, evaluated at $\theta = \pi/4$. Parameters are the same as in Fig. \ref{fig:SM_Fig1a}.}
    \label{fig:SM_Fig2a}
\end{figure}
\begin{figure}[h!]
    \centering
    \includegraphics[width=0.65\columnwidth]{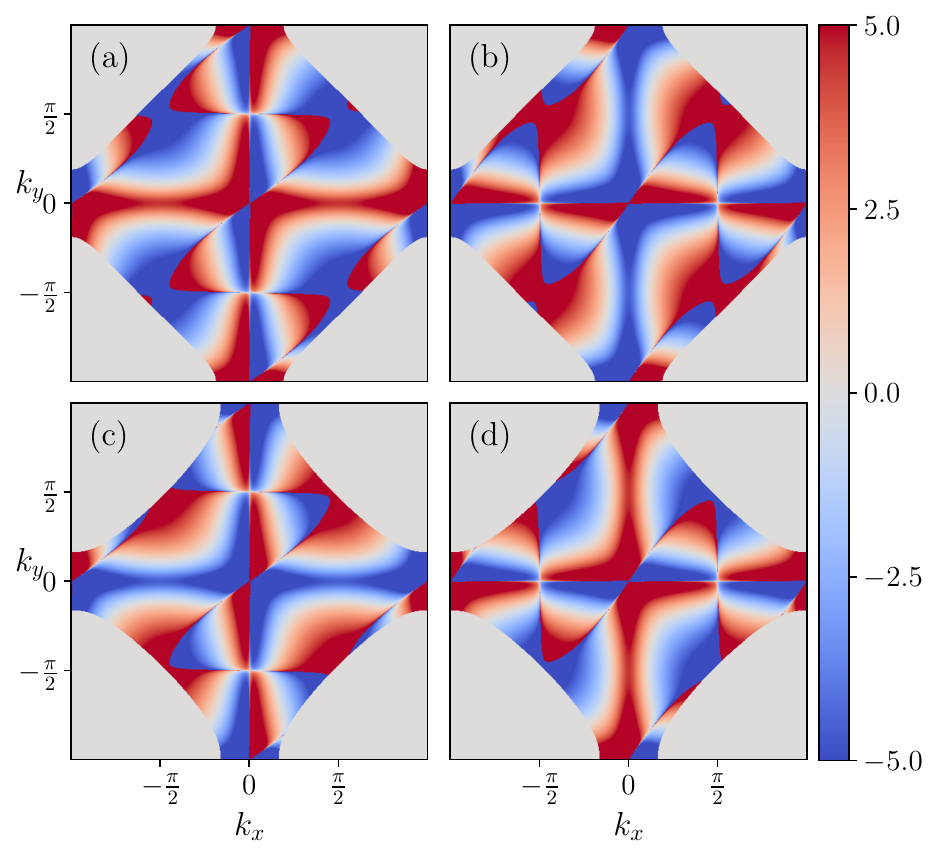} 
    \caption{Momentum distribution of the Fermi-function-weighted derivatives of the field-induced Berry curvature in log-scale for the two bands, (a) $f_0~\partial_{k_x}\Omega^{\mathrm{E}}_1$, (b) $f_0~\partial_{k_y}\Omega^{\mathrm{E}}_1$, and for band 2, (c) $f_0~\partial_{k_x}\Omega^{\mathrm{E}}_2$, (d) $f_0~\partial_{k_y}\Omega^{\mathrm{E}}_2$, evaluated at $\theta = \pi/4$. Parameters are the same as in Fig. \ref{fig:SM_Fig1b}.}
    \label{fig:SM_Fig2b}
\end{figure}
\begin{figure}[h!]
    \centering
    \includegraphics[width=\columnwidth]{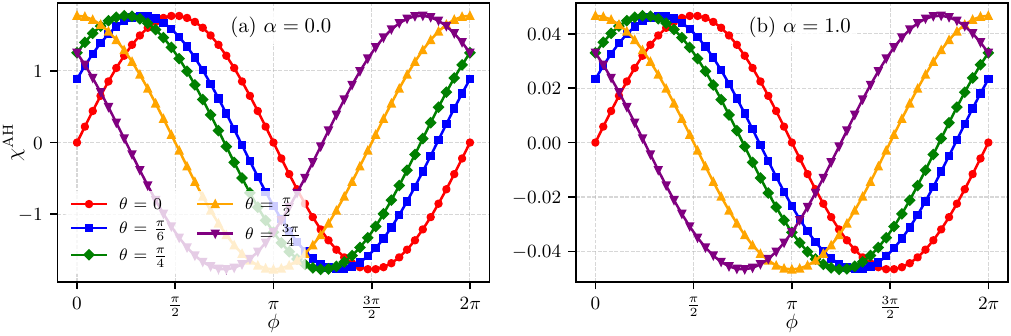}
     \caption{Second-order Hall conductivity $\chi^{\mathrm{AH}}$ as a function of angle $\phi$, set by the ac driving field $\mathbf{E}^\omega$, for
several different angles $\theta$, set by the symmetry-breaking static field $\mathbf{\mathbf{E}}^{\mathrm{dc}}$, with the $x$-axis for (a) $\alpha=0$, (b) $\alpha=1$. $\chi^{\mathrm{AH}}$ is scaled by the prefactor $\mathcal{A}=\frac{e^3 \tau}{2 (1 + i \omega \tau) \hbar^2} $  and the parameters are the same as in Fig.~\ref{fig:SM_FigBS}.}
   \label{fig:SM_FigS3}
\end{figure}

\section{Topography of Fermi surfaces}
\begin{figure}[H]
    \centering
    \includegraphics[width=\columnwidth]{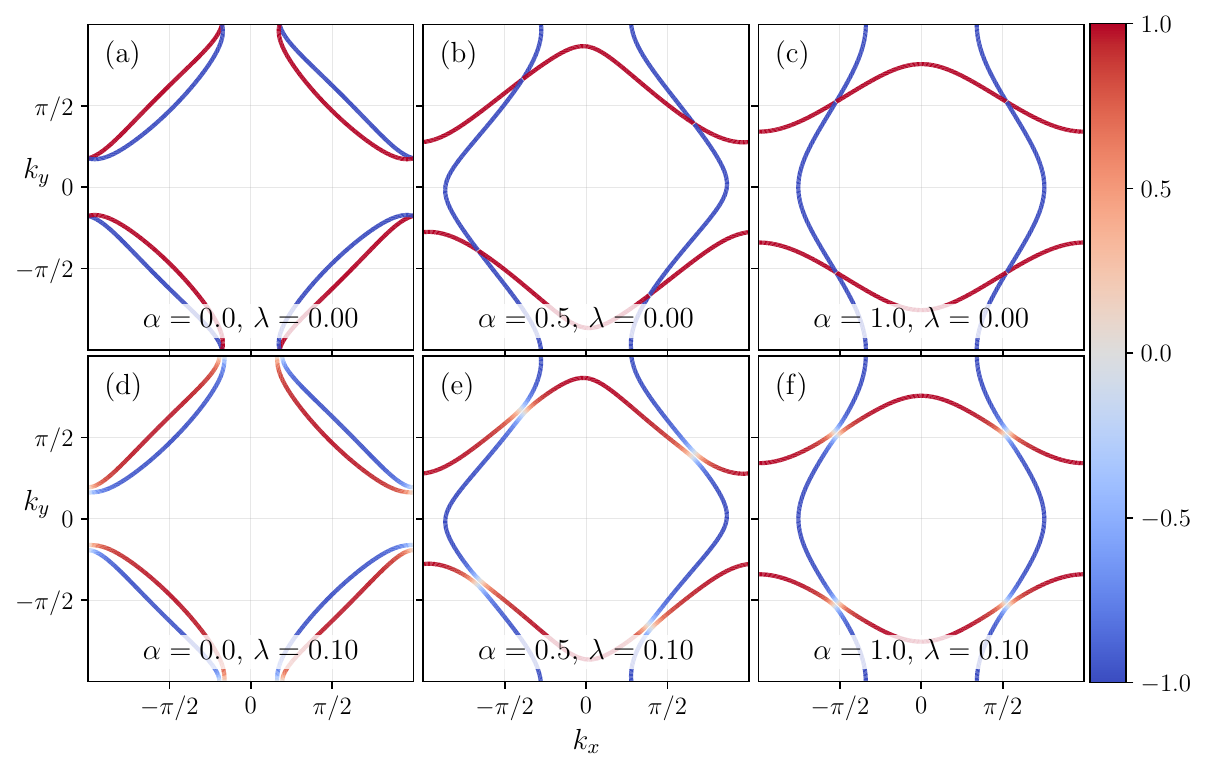}
   \caption{Spin-split Fermi surfaces (a,d) $\alpha=0$; (b,e) $\alpha=0.5$ and (c,f) $\alpha=1.0$. Upper row is for $\lambda=0$ and lower row for $\lambda=0.1 t$. Chemical potential is set to $\mu=0.3t$, same as the main text. }
    \label{fig:SM_FigS4a}
    \end{figure}

\begin{figure}[H]
    \centering
    \includegraphics[width=\columnwidth]{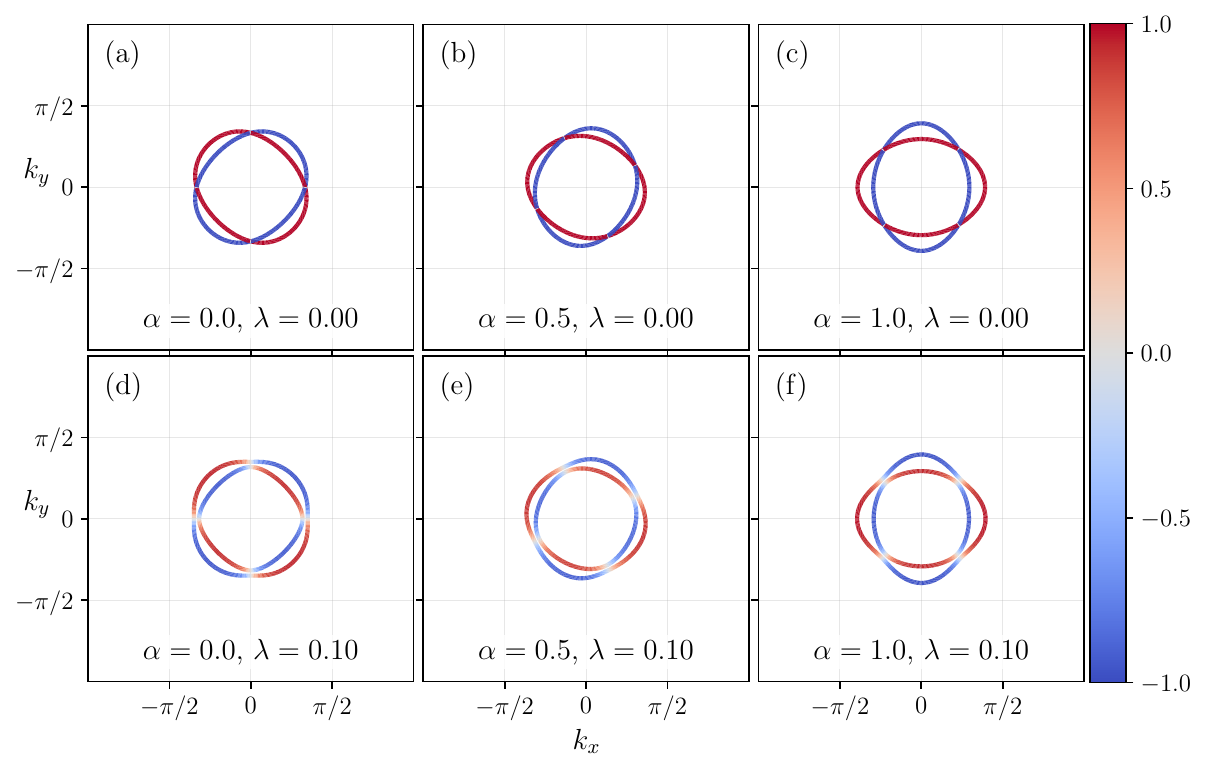}
     \caption{Spin-split Fermi surfaces (a,d) $\alpha=0$; (b,e) $\alpha=0.5$ and (c,f) $\alpha=1.0$. Upper row is for $\lambda=0$ and  lower row for $\lambda=0.1 t$. Chemical potential is set to $\mu=-3.0 t$.}
    \label{fig:SM_FigS4b}
\end{figure}

\begin{figure}[H]
    \centering
    \includegraphics[width=\columnwidth]{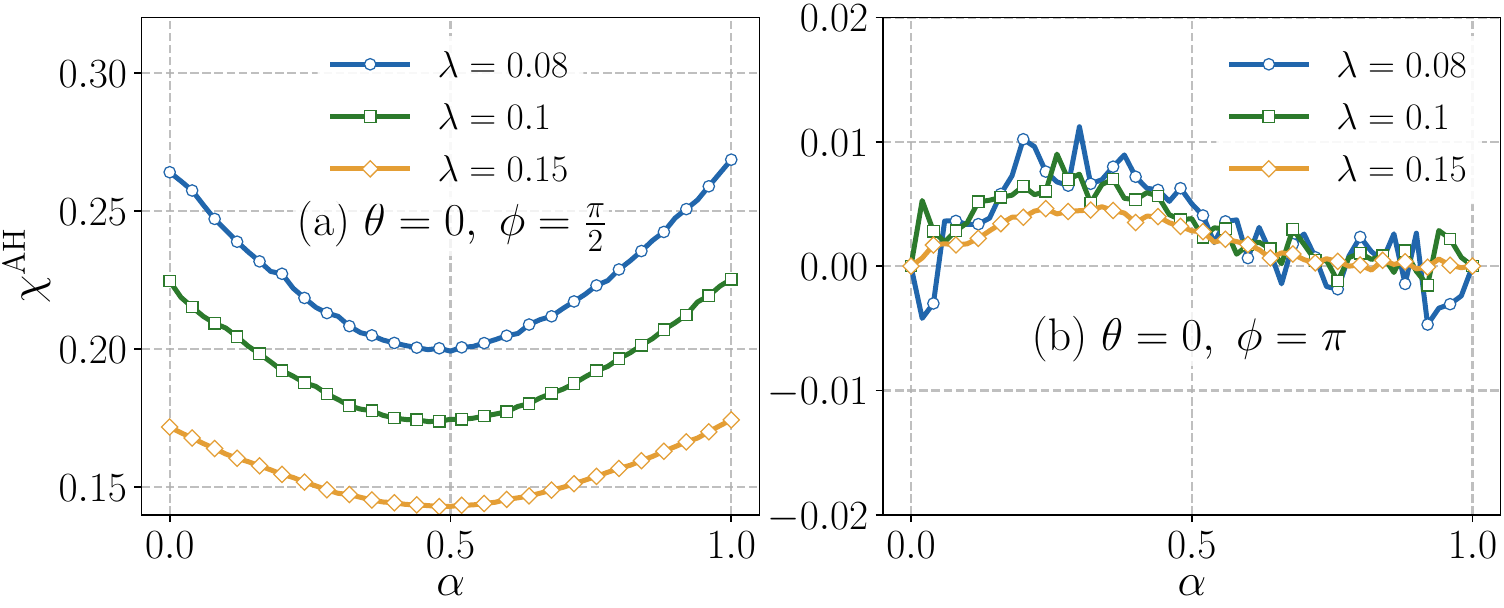}
\caption{Second-order Hall conductivity $\chi^{\mathrm{AH}}$ as a function of $\alpha$ for different RSOC $\lambda$ values and field orientations: (a) $\theta=0$, $\phi=\pi/2$ and (b) $\theta=0$, $\phi=\pi$ with $t = 1$, $t_{\text{am}}=0.5 t$, and $\mu=-3t$ (same as Fig.~\ref{fig:SM_FigBS}). }
    \label{fig:SM_FigS5}
    \end{figure}

In Fig.~\ref{fig:SM_FigS4a} and Fig.~\ref{fig:SM_FigS4b}, we present the spin-split Fermi surfaces for various altermagnetic orders across different doping regimes. In the absence of Rashba spin–orbit coupling (RSOC) $\lambda$, band-touching points are present, but a gap opens at these points as soon as $\lambda$ is tuned to a finite value. The second-order Hall response increases with decreasing RSOC $\lambda$, as shown in Fig.~\ref{fig:SM_FigS5}. This is because for a smaller value of $\lambda$, the energy gaps between bands are reduced, leading to an enhanced quantum metric or Berry curvature polarization (BCP), which in turn results in a stronger field-induced second-order response.
In Fig.~\ref{fig:SM_FigS5}(a), we show the variation of $\chi^\text{AH}$ with $\alpha$ for $\mu=-3 t$ keeping $\mathbf{E}^\text{dc}\perp \mathbf{E}^\omega$. In contrast to the results in Fig.~\ref{fig4} in the main text, we find that $\chi^\text{AH}$ is finite similar in magnitude for all $\alpha$. In this doping limit, there are also two electron pockets for all $\alpha$ in contrast to the doping limit considered in the main text. In this case, the contribution of the two similar Fermi surfaces add constructively to give an overall finite response for any $\alpha$. In Fig.~\ref{fig:SM_FigS5}(b), the same plot is shown for $\mathbf{E}^\text{dc}\parallel \mathbf{E}^\omega$, with an overall small $\chi^\text{AH}$, showing that also in this case there is large tunability with field-directions.


\end{document}